\documentclass[letterpaper,times]{IONconf}

\usepackage{times}
\usepackage{epsfig}
\usepackage[T1]{fontenc}
\usepackage{graphicx}
\usepackage{booktabs}
\usepackage[misc]{ifsym}

\usepackage{paralist}
\usepackage{eurosym}
\usepackage{flushend}
\usepackage{amsmath}
\usepackage{amssymb}
\usepackage{url}
\usepackage[utf8]{inputenc}
\usepackage{amsfonts}
\usepackage{soul}
\usepackage[justification=centering]{caption}
\usepackage[list=true]{subcaption}
\usepackage{tabu}
\usepackage{wrapfig}
\usepackage{arydshln}
\usepackage[normalem]{ulem}
\usepackage{multirow}
\usepackage{siunitx}
\usepackage{natbib}
\usepackage{hyperref}
\usepackage{makecell}
\usepackage{adjustbox}
\usepackage{array}
\usepackage{diagbox}
\usepackage{tikz}
\usetikzlibrary{arrows.meta,positioning,calc,fit,backgrounds,shapes.geometric,shadows}

\title{Suppressing Multiple GNSS Interferers with Local-Gradient Adaptive Notch Filtering}

\author{Burak Soner\textsuperscript{1}, Felix Ott\textsuperscript{2}\\
\textsuperscript{1}\textit{sobu Labs, Ankara, Türkiye}\\
\vspace{0.1cm}
\textsuperscript{2}\textit{Fraunhofer Institute for Integrated Circuits IIS, Nürnberg, Germany}\\
{\tt\small burak@sobulabs.com, felix.ott@iis.fraunhofer.de}
}

\begin{document}

\maketitle

\section*{BIOGRAPHY}

\biography{Burak Soner}{received his B.Sc.~in mechatronics from Sabancı University and his Ph.D.~in electrical and electronics engineering from Koç University, both in Istanbul, Türkiye. He is a multidisciplinary engineer with more than ten years of simultaneous research and professional experience in power electronics, control, wireless communication and sensing, optics, embedded systems, FPGAs, and artificial intelligence (AI) for edge devices. He recently founded his own company, \emph{sobu Labs}, in Ankara, Türkiye, providing research \& engineering services for compact DSP / AI algorithms on FPGAs and other SoCs, primarily for autonomous navigation applications, for customers in Türkiye, EU and USA.}

\biography{Felix Ott}{received his M.Sc.~degree in Computational Engineering from Friedrich-Alexander University (FAU) Erlangen-Nürnberg in 2019. He subsequently joined the Self-Learning Systems group within the Machine Intelligence department at the Fraunhofer Institute for Integrated Circuits IIS. In 2023, he was awarded his Ph.D.~from Ludwig-Maximilians University (LMU) Munich, where he conducted research in the Probabilistic Machine and Deep Learning group. He is currently serving as a postdoctoral researcher and manages the program Signals Intelligence (SIGINT). His research focuses on representation learning, domain adaptation, and few-shot learning for GNSS-based signals interference monitoring, such as interference signal classification, characterization, direction and source localization finding, and mitigation.}

\section*{ABSTRACT}

Adaptive notch filters (ANFs) based on the normalized least mean squares (NLMS) algorithm are an established technique for suppressing narrowband interference in GNSS receivers, especially chirp-based personal privacy devices (PPDs). The traditional solution for multiple interferers is simply cascading the NLMS-ANF into multiple stages but this solution degrades performance when multiple chirp jammers of similar power are present. With multiple comparable chirps, competing full-band gradient contributions can partially cancel and drive the notch between interferers, resulting in ineffective suppression. We propose a low-complexity modification to the cascaded NLMS-ANF update rule to resolve this: a local gradient (LG) formulation -- LG-ANF.  The LG-ANF forms a locally bandlimited gradient signal by translating the input to baseband around the current frequency estimate, low-pass filtering it, and translating it back, thereby attenuating competing interferers before adaptation. The modification generalizes the traditional NLMS-ANF to handle the similar-power multiple interferer problem, using just one additional parameter (i.e., the low-pass bandwidth, which collapses the LG-ANF to the NLMS-ANF formulation at its maximum value of $1.0$) with negligible computational load. We characterize the gradient-averaging failure analytically and visually, demonstrate independent chirp tracking in the worst-case two-chirp scenarios, and evaluate statistical suppression performance by sweeping amplitude ratio against a standard NLMS-ANF, a frequency-locked-loop (FLL)-based ANF and a per-sample oracle upper bound. We further characterize the LG advantage over various ANF parameters and chirp conditions. Simulations show that the LG-ANF reduces to the standard NLMS-ANF outside the pathological multi-chirp regime, while recovering up to $4$\,dB of in-band suppression around pathological amplitude ratios under per-condition optimization of the local bandwidth, and up to $13$\,dB for selected edge-case parameter combinations. These optimized results represent an upper bound on the achievable LG gain. In a hardware experiment using a fixed $\alpha_{lp}=0.5$, the LG-ANF extends position-lock survival by $24$\,s relative to the standard ANF.

\section{INTRODUCTION}
\label{label_introduction}

GNSS receivers are highly vulnerable to radio-frequency interference because satellite signals arrive with very low power. A personal privacy device (PPD) often transmits sawtooth chirps across GNSS frequency bands. Such chirps sweep their instantaneous frequency over a fixed period and then reset, producing wideband, time-varying interference. While single chirp jammers are already challenging, multiple simultaneous chirps can arise from nearby independent PPDs or from one jammer transmitting several components, making multi-interferer mitigation a practically relevant scenario.

Adaptive notch filtering is a classical low-complexity solution for suppressing such GNSS interferers. The minimal-parameter ANF with constrained poles and zeros was introduced by~\cite{nehorai1985} and analyzed by~\cite{stoica1988}, with later improvements to the filter structure including lattice-based and high-order structures~\citep{regalia1991,kim2014}, and zero-phase filtering~\citep{khan2024anf_icm}. In GNSS, notch filters have been widely used as front-end mitigation tools due to their modest computational cost~\citep{borio2006}. However, inaccurate or overly aggressive notch placement can distort the useful GNSS signal through effects such as group delay and timing bias~\citep{borio2020}, making reliable frequency tracking and experimental validation of the filtered signal with a GNSS receiver chain essential.

For swept chirp interference, the notch center must follow a rapidly varying instantaneous frequency. Prior work has considered FLL-based ANFs for mitigating fast sawtooth chirp jammers with second-order control loops~\citep{gamba2018}, multi-parameter ANFs for additionally handling the residual energy that leaks at the chirp resets~\citep{vandermerwe2023}, spectral-NLMS hybrid approaches for improving the resolution-bandwidth trade-off~\citep{khan2024hybrid}, efficient cascaded implementations~\citep{song2024multiplierfree}, and classification-driven suppression with the optimal mitigation technique selection~\citep{deng2025differentiated}. Noting that the gradient-based methods for ANFs are actually fully stochastic (per sample update) ``online learners'' of the instantaneous chirp jammer frequency~\citep{soner2026adam}, machine learning (ML) methods in the GNSS interference mitigation space should also be mentioned. Interference classification~\citep{ott_heublein_jispin,heublein_feigl_crpa}, characterization~\citep{khan_raichur,heublein_feigl_jispin}, mitigation selection, array anti-jamming, and ANF parameterization~\citep{chen2025gnss_dl_mitigation,xie2025dnn_antijamming,ghanbarzade2025gnssgps,jiang2025acsnet,chaiben2025chirp_cnn,abbasi2020anf_ann,tahouri2021eanf,austermann2023ml_anf} have recently been studied in this direction. These approaches broaden the design space, but they typically introduce additional estimation, classification, training, or optimization stages.

The standard approach for multiple interferers is to cascade several ANF stages (typically NLMS-based, or a variant of it), where each ANF is expected to track and suppress one interferer from the residual of the previous stage. This approach can fail when multiple chirps of comparable power coexist. The reason is that the NLMS update computes its gradient from the full-band input. Consequently, the gradient contains competing contributions from all chirps, each pulling the notch toward its own instantaneous frequency. When these contributions have similar magnitude, they can cancel, causing the notch to settle between chirps and track neither. Such degradation of cascaded ANFs in complex multi-interferer scenarios has been observed empirically~\citep{vandermerwe2023_multipleinterferer}. Importantly, this is not a general limitation of the NLMS-ANF across all multi-interferer conditions, but a localized failure regime that occurs when competing gradient contributions become sufficiently balanced. The objective of the proposed LG-ANF is therefore not to outperform the standard ANF universally, but to avoid this pathological regime while retaining conventional NLMS behavior when local gradient conditioning is unnecessary.

We address this failure with a modification called the \textit{local-gradient ANF} (LG-ANF). Instead of changing the notch structure, objective, or NLMS update form, we change the signal used to compute the gradient. Considering the online learner interpretation of the NLMS-ANF, this modification is conceptually similar to task-specific input conditioning used in learning systems: rather than changing the model or objective, we parametrically modify the input data so that the gradient is dominated by the component relevant to the current task (mitigating individual chirps). Specifically, the input is translated to baseband relative to the current frequency estimate, low-pass filtered, and translated back. The resulting locally bandlimited signal suppresses competing chirps outside the neighborhood of the current estimate before gradient computation, while the actual notch-filter data path still operates on the full input. This adds only one first-order low-pass filter and two complex mixings per stage, controlled by a single bandwidth parameter, and collapses to the standard NLMS-ANF at maximum bandwidth.

\textbf{Contributions.} The contributions of this paper are as follows: (i) an analysis of the gradient-averaging mechanism that causes cascaded ANFs to fail under multiple comparable chirp interferers, (ii) the LG-ANF update, which locally decouples competing gradient contributions in the pathological multi-chirp regime while recovering the conventional NLMS update at maximum local bandwidth, and (iii) simulation and experimental evaluation with a GNSS-SDR showing improved multi-chirp suppression and extended position-lock survival. (iv) The corresponding recordings and metadata for the experiments are released for reproducibility~\citep{soner2026_dataset}.

\textbf{Outlook.} The remainder of this paper is organized as follows. Section~\ref{label_related_work} reviews related work on ANFs, GNSS interference mitigation, and learning-based approaches. Section~\ref{label_methodology} introduces the considered multi-chirp interference problem, analyzes the gradient-averaging mechanism, and presents the proposed local-gradient ANF. Section~\ref{label_simulation} evaluates the method in controlled simulations, including chirp-tracking experiments, amplitude-ratio sweeps, and parameter-map analyses. Section~\ref{label_real_world_experiments} validates the proposed method on recorded hardware data using a GNSS-SDR and an independent position reference. Finally, Section~\ref{label_conclusion} concludes the paper.
\section{RELATED WORK}
\label{label_related_work}

\subsection{Adaptive Notch Filters (ANFs)}
\label{label_related_work_ANF}

\cite{nehorai1985} introduced the minimal-parameter ANF with constrained poles and zeros, forming the basis of the first-order infinite impulse response (IIR) ANF, and \cite{stoica1988} analyzed the performance of such constrained-pole-zero ANFs. Their analyses clarify classical ANF behavior but does not treat multi-chirp gradient cancellation. Our work extends this perspective by analyzing the two-chirp gradient decomposition and correcting it through local bandlimiting. \cite{regalia1991,kim2014} proposed improved lattice-based IIR ANFs of varying orders with better implementation robustness. These change the filter realization but do not address the multi-interferer issue. Our method instead keeps the standard lightweight ANF structure intact and modifies only the gradient input, with minimal additional complexity. \cite{borio2006} analyzed one-pole notch filtering for GNSS interference mitigation and quantified associated suppression and loss behavior. The work establishes notch filtering for GNSS but the scope is limited to classical single-source narrowband settings. Our method focuses on the failure of cascaded ANFs under multiple overlapping chirp interferers. \cite{gamba2018} studied FLL-based ANFs for tracking swept jammers. While FLLs are effective for single chirps, their discriminators are even worse than NLMS at distinguishing multiple phase components. Our LG-ANF is gradient-based and reduces this negative effect by computing the update from a local spectral neighborhood. \cite{borio2020} showed that interference mitigation can affect GNSS timing through effects such as group delay and timing bias. This motivates avoiding unnecessarily aggressive or misplaced notches. \cite{vandermerwe2023_multipleinterferer} showed empirically that cascaded ANFs can degrade in complex multi-interferer scenarios. The work identifies the problem but does not provide a minimal correction to the NLMS gradient. Our paper explains the failure through gradient averaging and proposes such a correction. \cite{vandermerwe2023} proposed MPANF, which adapts multiple notch parameters for improved mitigation and adds a pulse blanker to handle the fast reset of sawtooth chirps. This increases flexibility but also adds parameter and stability complexity, and relies on FLLs, which are disadvantaged in multiple interferer settings. Our approach instead keeps the single-parameter NLMS-ANF and improves only the gradient computation to handle multiple interferers.

\subsection{Machine Learning in Interference Mitigation}
\label{label_related_work_ML_ANF}

Viewed from an adaptation perspective, gradient-based ANFs can be interpreted as stochastic, per-sample online estimators of the instantaneous jammer frequency. This connection motivates a brief discussion of ML-based GNSS interference mitigation methods, especially those that use learned models or optimization-inspired updates to cope with diverse and time-varying jamming conditions. ML has recently been explored for GNSS interference detection, classification, and mitigation, mainly as a way to handle diverse and time-varying jamming conditions~\citep{heublein_feigl_crpa}. \cite{chen2025gnss_dl_mitigation} proposed a deep-learning-based interference mitigation framework that selects a suitable suppression strategy according to the observed interference characteristics. However, the method acts as a high-level selection mechanism rather than improving the internal adaptation rule of a specific filter. \cite{xie2025dnn_antijamming} used an LSTM for GNSS array anti-jamming, showing that learned models can exploit nonlinear structure in multi-antenna data, but this approach requires array observations and is therefore different from our single-channel pre-correlation ANF setting. \cite{ghanbarzade2025gnssgps} studied ML-methods for GNSS jamming and spoofing identification, but their focus is primarily on detection and classification rather than continuous front-end interference suppression. \cite{jiang2025acsnet} proposed ACSNet for compound GNSS jamming classification using a convolutional architecture, which is useful for interference awareness but does not directly remove the jammer from the received signal. \cite{chaiben2025chirp_cnn} investigated CNN-based classification of FM chirp jammers from time-frequency representations, whereas our method does not require a trained classifier and instead modifies the ANF update rule itself.

Learning-based ideas have also appeared in connection with ANF, although this literature is more limited. \cite{khan2024anf_icm} combined ANF-based interference characterization with an ML-classifier to distinguish interference types and select mitigation behavior, but the ANF itself still suffers from the usual multi-component adaptation limitations. \cite{abbasi2020anf_ann} combined an ANF with an ML-model for GPS continuous-wave jamming cancellation, which introduces a learned correction mechanism but targets tone-like interference rather than overlapping chirp jammers. \cite{tahouri2021eanf} designed an evolutionary ANF for GPS anti-jamming using genetic and particle-swarm optimization, but such evolutionary search increases optimization complexity and is not intended as a simple online NLMS update. \cite{austermann2023ml_anf} used ML to parameterize ANFs for EMI noise reduction, which is conceptually related to learned ANF tuning but outside the GNSS chirp-mitigation setting. \cite{soner2026adam} proposed an Adam-inspired ANF update for faster frequency tracking, demonstrating that ML-optimizers can improve ANF adaptation. Whereas, our LG-ANF keeps the NLMS update structure and instead changes only the signal from which the gradient is computed. Thus, existing ML-based approaches either classify interference, select mitigation strategies, tune ANF parameters, or replace the optimizer, while the proposed LG-ANF directly addresses the multi-chirp gradient-averaging failure with a lightweight and simple modification of the gradient computation.

\section{PROPOSED METHOD}
\label{label_methodology}

\begin{figure*}[!t]
\centering
\resizebox{\textwidth}{!}{%
\begin{tikzpicture}[
    font=\scriptsize,
    >=Latex,
    box/.style={
        rectangle,
        draw,
        rounded corners,
        align=center,
        minimum height=0.72cm,
        minimum width=1.95cm,
        inner sep=2.2pt,
        fill=blue!7
    },
    data/.style={
        rectangle,
        draw,
        rounded corners,
        align=center,
        minimum height=0.72cm,
        minimum width=2.05cm,
        inner sep=2.2pt,
        fill=green!8
    },
    grad/.style={
        rectangle,
        draw,
        rounded corners,
        align=center,
        minimum height=0.72cm,
        minimum width=1.85cm,
        inner sep=2.2pt,
        fill=orange!10
    },
    warn/.style={
        rectangle,
        draw,
        rounded corners,
        align=center,
        minimum height=0.72cm,
        minimum width=2.15cm,
        inner sep=2.2pt,
        fill=red!8
    },
    sum/.style={
        circle,
        draw,
        align=center,
        minimum size=0.45cm,
        inner sep=1pt,
        fill=gray!8
    },
    arrow/.style={->, thick},
    dashedarrow/.style={->, thick, dashed},
    group/.style={
        rectangle,
        draw,
        dashed,
        rounded corners,
        inner sep=0.22cm
    }
]


\node[box] (c1) at (0,0.45) {Chirp 1\\$f_1[n],A_1,T_{c,1}$};
\node[box] (c2) at (0,-0.45) {Chirp 2\\$f_2[n],A_2,T_{c,2}$};

\node[sum] (plus) at (1.5,0) {$+$};

\node[box, minimum width=2.75cm] (rx) at (3.9,0)
{Received signal\\$x[n]=x_1[n]+x_2[n]+w[n]$};

\node[warn, minimum width=2.35cm] (std) at (7.55,0)
{Standard cascaded ANF\\full-band gradient};

\node[warn, minimum width=2.75cm] (pulls) at (10.75,0)
{Competing terms\\$g_1[n]+g_2[n]+g_{\times}[n]$};

\node[warn, minimum width=2.65cm] (bad) at (14.15,0)
{Gradient averaging\\tracks neither chirp};

\draw[arrow] (c1.east) -- (plus);
\draw[arrow] (c2.east) -- (plus);
\draw[arrow] (plus) -- (rx);
\draw[arrow] (rx) -- (std);
\draw[arrow] (std) -- (pulls);
\draw[arrow] (pulls) -- (bad);

\node[group,
      fit=(c1)(c2)(plus)(rx)(std)(pulls)(bad),
      label={[font=\bfseries\small]above:Multi-Chirp Problem in Cascaded ANFs}] (topgroup) {};


\node[box] (xin) at (0,-2.4) {Full input\\$x[n]$};

\node[data, minimum width=2.35cm] (anf) at (3.0,-2.4)
{Full-band ANF\\data path};

\node[data, minimum width=1.95cm] (yout) at (5.95,-2.4)
{Output\\$y[n]$};

\node[data, minimum width=2.15cm] (cascade) at (8.8,-2.4)
{Cascade\\$K$ stages};

\draw[arrow] (xin) -- (anf);
\draw[arrow] (anf) -- (yout);
\draw[arrow] (yout) -- (cascade);

\node[grad] (mixdown) at (0,-4.0)
{Mix down\\$e^{-j\theta[n]}$};

\node[grad] (lpf) at (2.8,-4.0)
{Local LPF\\$\alpha_{lp}$};

\node[grad] (mixup) at (5.6,-4.0)
{Mix up\\$e^{j\theta[n]}$};

\node[grad, minimum width=2.25cm] (locstate) at (8.6,-4.0)
{Local state\\$x_{r,\mathrm{loc}},y_{\mathrm{loc}}$};

\node[grad] (lg) at (11.6,-4.0)
{Local gradient\\$g_{\mathrm{LG}}[n]$};

\node[data, minimum width=2.05cm] (update) at (14.4,-4.0)
{Update\\$\theta[n+1]$};

\draw[arrow] (mixdown) -- (lpf);
\draw[arrow] (lpf) -- (mixup);
\draw[arrow] (mixup) -- (locstate);
\draw[arrow] (locstate) -- (lg);
\draw[arrow] (lg) -- (update);

\draw[arrow] (xin.south) -- (mixdown.north);

\coordinate (thetaTop) at (14.4,-3.3);
\draw[dashedarrow] (update.north) -- (thetaTop);
\draw[dashedarrow] (thetaTop) -| node[pos=0.25, above, font=\scriptsize] {$\theta[n]$} (anf.south);

\node[data, minimum width=8.15cm, minimum height=0.4cm] (note) at (5.4,-5.25)
{Local bandlimiting attenuates competing chirps before gradient computation; the nearest chirp self-term dominates the update.};

\draw[dashedarrow] (lpf.south) -- ++(0,-0.35) -| (note.north);

\node[group,
      fit=(xin)(anf)(yout)(cascade)(mixdown)(lpf)(mixup)(locstate)(lg)(update)(note),
      label={[font=\bfseries\small]above:Proposed Local-Gradient Cascaded ANF}] (botgroup) {};

\end{tikzpicture}%
}
\caption{Overview of the multi-chirp gradient-averaging problem and the proposed local-gradient ANF (LG-ANF). In the standard cascaded NLMS-ANF, the gradient is computed from the full-band input and contains competing contributions from all chirp interferers. The proposed LG-ANF keeps the full-band ANF data path unchanged, but computes the adaptation gradient from a locally bandlimited signal obtained by mixing around the current frequency estimate, low-pass filtering, and mixing back.}
\label{figure_method_overview}
\end{figure*}
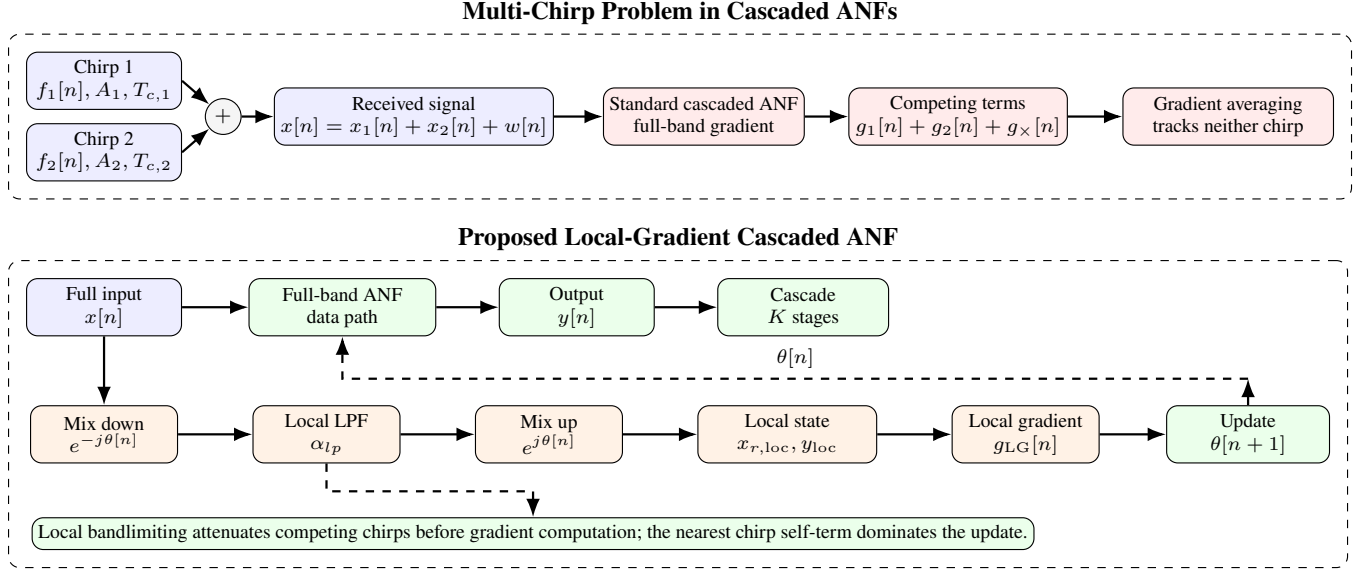

The method is structured around the central failure mode of cascaded ANFs and the proposed local-gradient correction, as summarized in Figure~\ref{figure_method_overview}. We first define the multi-chirp GNSS interference scenario in Section~\ref{label_methodology_problem}. Section~\ref{label_methodology_ANF} then reviews the first-order complex NLMS adaptive notch filter, which forms the baseline of the proposed method. In Section~\ref{label_methodology_gradient_averaging}, we analyze why the standard cascaded ANF fails when several chirps of comparable power coexist: the full-band gradient contains competing contributions from all interferers, which can cancel and drive the notch toward a frequency occupied by none of them. Finally, Section~\ref{label_methodology_local_gradient} introduces the proposed LG-ANF, which computes the adaptation gradient from a locally bandlimited version of the input signal, thereby attenuating competing chirps before the update is formed.

\subsection{Problem Definition}
\label{label_methodology_problem}

PPDs are low-cost, widely available jammers that transmit sawtooth chirp signals in GNSS frequency bands to disable the self-localization capability of nearby receivers. A single sawtooth chirp sweeps instantaneous frequency linearly from $f_0$ to $f_1$ over a period $T_c$, then resets, producing wideband, time-varying interference. As PPD usage grows, the simultaneous presence of two or more independent chirp interferers, each with its own sweep rate, frequency range, and amplitude, becomes a practically relevant and understudied threat scenario. Moreover, transmitting multiple chirps at the same time from the same unit is also a practical possibility, further increasing the risk of this threat scenario. We argue that the classical cascaded NLMS-ANF structure that uses the full-band input signal is disadvantaged in such multiple-interferer cases. We thus theoretically demonstrate this, and propose a solution at improving mitigation performance in the following subsections.

\subsection{The Classical NLMS Adaptive Notch Filter}
\label{label_methodology_ANF}

The first-order complex NLMS-ANF~\citep{nehorai1985} suppresses a single chirp by tracking its instantaneous frequency with a single scalar state $\theta[n]$. Defining $z_0 = e^{j\theta[n]}$, the filter operates as:
\begin{align}
x_r[n] &= x[n] + k_a\, z_0\, x_r[n-1], \label{eq:res}\\
y[n]   &= x_r[n] - z_0\, x_r[n-1], \label{eq:out}
\end{align}
where $x[n]$ is the input, $y[n]$ is the notch-filtered output, $x_r[n]$ is the filter state, and $k_a \in (0,1)$ controls notch width. The frequency estimate is updated each sample by gradient descent on the output power $|y[n]|^2$:
\begin{equation}
\theta[n+1] = \theta[n] - \frac{\mu}{P[n] + \varepsilon}\, g[n], \label{eq:update}
\end{equation}
where
\begin{equation}
g[n] = 2\,\text{Re}\!\left\{y^*[n]\cdot (k_a - 1)\,j z_0\,x_r[n-1]\right\}
\end{equation}
is the stochastic gradient, i.e., the exact gradient of $|y[n]|^2$ w.r.t.~$\theta$ treating $x_r[n-1]$ as stationary, and $P[n] = \alpha_P |x_r[n]|^2 + (1-\alpha_P)P[n-1]$ is a running power estimate used for step size normalization. This simple adaptive filter is very popular in the GNSS interference mitigation field due to its high effectiveness~\citep{borio2006}, and for multiple interferers, the standard approach is cascading them in series. 

\subsection{The Gradient Averaging Problem}
\label{label_methodology_gradient_averaging}

The failure of cascaded ANFs in multi-interferer scenarios has been observed empirically~\citep{vandermerwe2023_multipleinterferer}, and the culprit is the gradient terms. The following analysis characterizes the underlying mechanism by deriving an approximate analytical expression for the gradient over an example of two coexisting chirp interferers. 

To derive the gradient independently of any particular adaptation trajectory, we analyze the batch objective $J_N(\theta) = \frac{1}{N}\sum_{n=0}^{N-1}|y_\theta[n]|^2$ rather than the stochastic objective defined above. We assume $\theta$ is fixed (or slowly varying) along the entire signal for tractable analysis. Unlike the stochastic gradient in Equ.~\ref{eq:update}, which treats $x_r[n-1]$ as a constant past state, the gradient of $J_N$ must account for how $\theta$ propagates through the full infinite impulse response (IIR) chain. Defining $\dot{z}_0 = jz_0$ as the derivative of $z_0$ with respect to $\theta$, and $s[n] \triangleq \partial x_r[n]/\partial\theta$ as the sensitivity state with $s[-1] = 0$, differentiating Equ.~\ref{eq:res} at fixed $\theta$ gives
\vspace{5px}
\begin{equation}
s[n] = k_a\!\left(\dot{z}_0\,x_r[n-1] + z_0\,s[n-1]\right),
\label{eq:sens}
\end{equation}
from which the exact output gradient of the filter output to $\theta$ in this formulation follows as
\vspace{5px}
\begin{equation}
\frac{\partial y[n]}{\partial\theta} = s[n] - \dot{z}_0\,x_r[n-1] - z_0\,s[n-1].
\label{eq:dydtheta}
\end{equation}
Note that this is a generalization of the stochastic gradient in Equ.~\ref{eq:update} as it can be recovered by freezing $s[n-1] = 0$ in Equ.~\ref{eq:sens}, which yields $\partial y/\partial\theta = (k_a-1)\dot{z}_0 x_r[n-1]$. Under the linearity assumption, we expand this formulation to see the individual contributing terms of the gradient. Specifically, when $x[n] = x_1[n] + x_2[n]$, the filter is linear in $x$ at fixed $\theta$, so $y = y_1 + y_2$ and $\partial y/\partial\theta = \partial y_1/\partial\theta + \partial y_2/\partial\theta$. The gradient of $J_N$ therefore splits exactly to three parts:
\begin{equation}
g[n] = \underbrace{g_1[n]}_{\text{self: chirp 1}} + \underbrace{g_2[n]}_{\text{self: chirp 2}} + \underbrace{g_{\times}[n]}_{\text{cross term}},
\label{eq:decomp}
\end{equation}
with $g_k[n] = 2\,\text{Re}\{y_k^*[n]\cdot \partial y_k/\partial\theta[n]\}$ and $g_{\times}[n] = 2\,\text{Re}\!\left\{y_1^*[n]\,\partial y_2/\partial\theta[n] + y_2^*[n]\,\partial y_1/\partial\theta[n]\right\}$.

\begin{figure}[!t]
    \centering
    \includegraphics[trim=10 10 10 10, clip, width=1.0\linewidth]{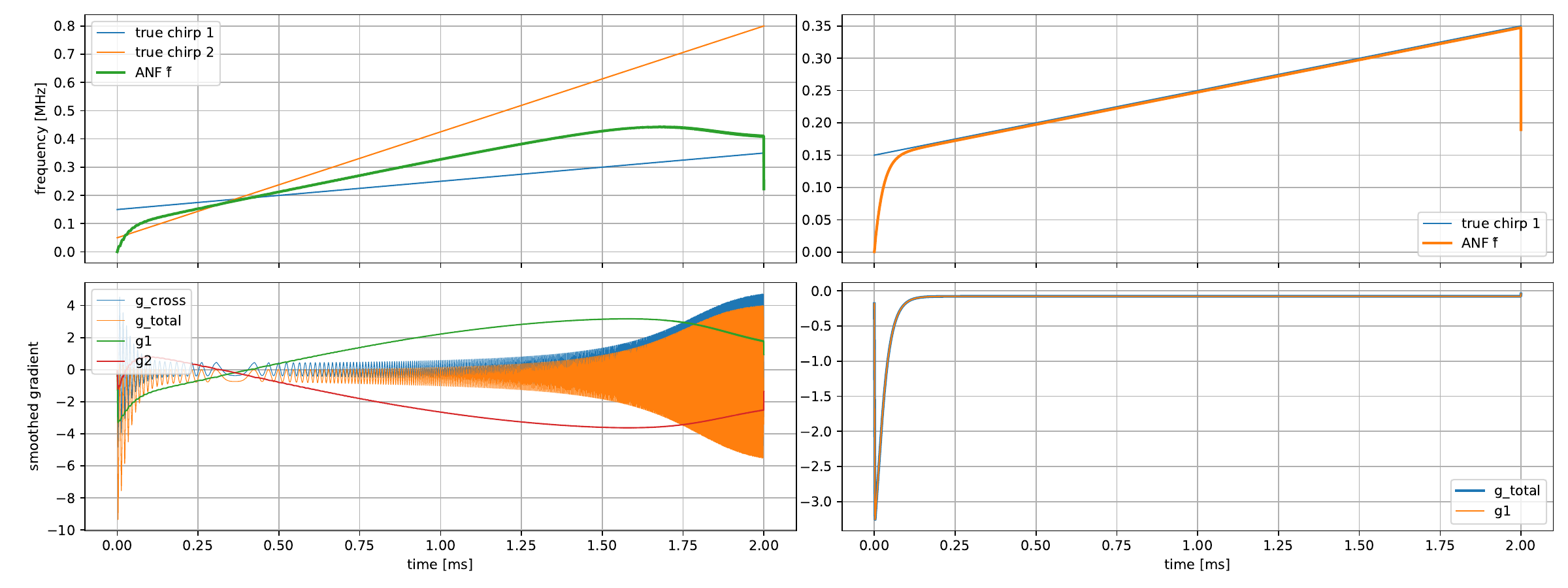}
    \caption{Gradient decomposition for double-chirp (left) and single-chirp (right). Top row: ANF frequency estimate vs.~true chirp frequencies. Bottom row: individual gradient components. In the double-chirp case, $g_1$ (green) and $g_2$ (red) carry opposite signs and practically cancel out; $g_\times$ (blue) oscillates with zero mean, leaving the total gradient $g_\text{total}$ (orange) without useful directional information.}
    \label{fig:grad}
\end{figure}

In the pathological case of $\theta$ being in the middle of the two chirps' instantaneous frequencies, $g_1$ and $g_2$ cancel each other out, and $g_\times$ oscillates rapidly at the beat frequency of the two chirps with zero mean. This makes the notch sweep a region occupied by neither jammer, suppressing useful GNSS signal while leaving both interferers intact. Figure~\ref{fig:grad} demonstrates a similar case in simulation where until approximately the 1.5\,ms mark, the ANF converges to neither chirp and instead tracks their mean.

\subsection{Local-Gradient Cascaded ANFs}
\label{label_methodology_local_gradient}

To work around the gradient averaging problem, we propose computing the gradient from a \textit{locally bandlimited} version of the input. The local signal $x_\text{loc}[n]$ is obtained by translating the input to baseband relative to the current estimate $\theta[n]$, applying a first-order IIR low-pass filter, and translating back:
\begin{align}
\tilde{x}[n]      &= x[n]\cdot e^{-j\theta[n]}, \label{eq:mix1}\\
\tilde{x}_{lp}[n] &= \alpha_{lp}\,\tilde{x}[n] + (1-\alpha_{lp})\,\tilde{x}_{lp}[n-1], \label{eq:lpf}\\
x_\text{loc}[n]   &= \tilde{x}_{lp}[n]\cdot e^{j\theta[n]}. \label{eq:mix2}
\end{align}
A separate filter driven by ~$x_\text{loc}$ yields $x_{r,loc}[n]$ and $y_\text{loc}[n]$ via recursions analogous to Equ.~\ref{eq:res}--\ref{eq:out}. Both the gradient and the power normalization use the local signal
\begin{equation}
g_\text{LG}[n] = 2\,\text{Re}\!\left\{y_\text{loc}^*[n]\cdot (k_a-1)\,j e^{j\theta}\,x_{r,loc}[n-1]\right\}.
\end{equation}
Since $x_\text{loc}$ retains only spectral content within a bandwidth proportional to $\alpha_{lp}$ around $\theta[n]$, the ``competing'' chirp is attenuated before gradient computation, and the filter ``snaps'' to the nearest chirp since the gradient is dominated by the self-term of the chirp nearest the current estimate. While this drives the frequency estimation of the given ANF stage, the data path of the interference mitigation method, i.e., Equ.~\ref{eq:res}--\ref{eq:out}, still operates on the full input $x[n]$, not the local version.

The modification introduces one new scalar parameter: $\alpha_{lp}$, the local bandwidth coefficient ($\alpha_P$, the power estimator time constant is already assumed to be present in the canonical NLMS-ANF solution). For $K$ simultaneous chirps, $K$ LG-ANF stages are cascaded.

\section{SIMULATIONS}
\label{label_simulation}

\subsection{Equal-Amplitude Chirp Tracking}
\label{label_simulation_chirp_tracking}

\begin{figure}[!h]
\centering
\vspace{-0.05cm}
\includegraphics[width=0.95\columnwidth]{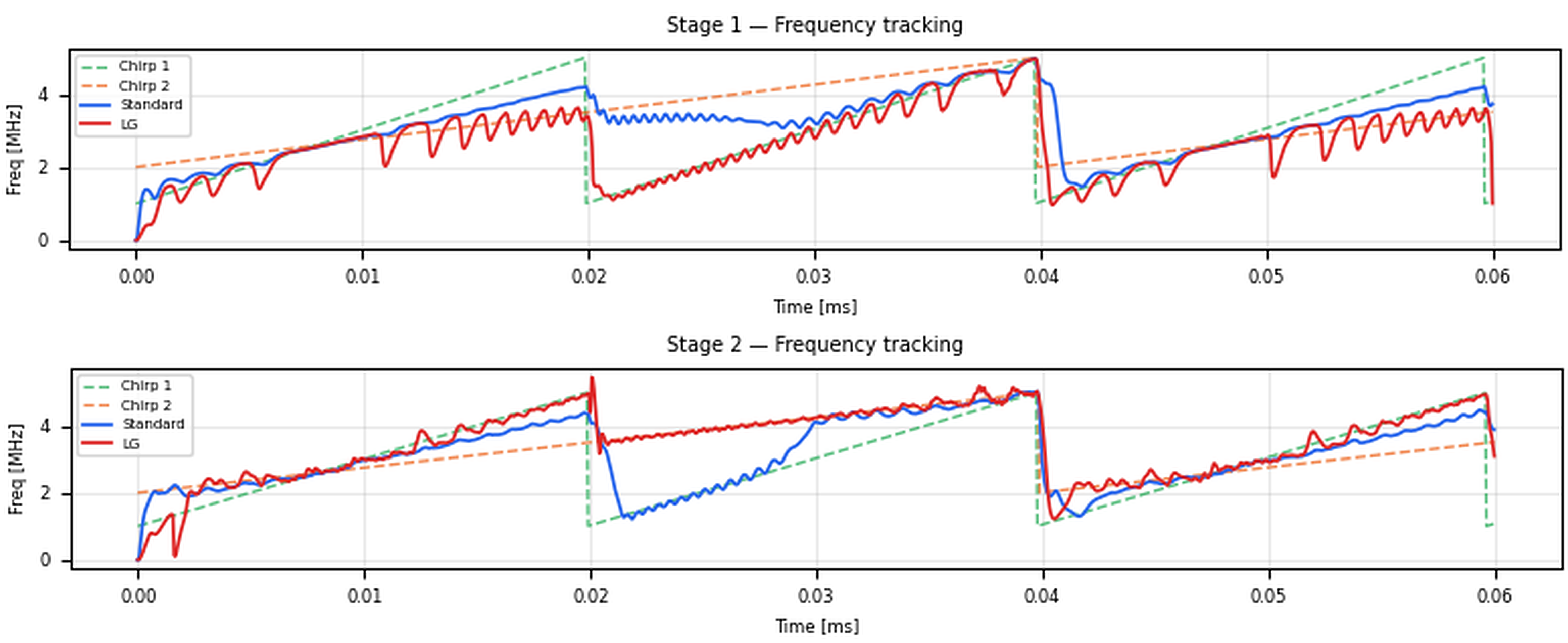}
\vspace{-0.35cm}
\caption{Illustrative frequency-tracking example to expose the gradient-averaging failure mode: Standard ANF (blue) vs.\ LG-ANF (red) in a two-stage cascaded architecture applied to two simultaneous equal-amplitude chirps. Dashed lines show true instantaneous chirp frequencies. The standard ANF oscillates between both chirps in both stages depending on the position of the two sweeps. The LG-ANF converges to and tracks each chirp independently.}
\label{fig:tracking}
\end{figure}

We initially simulate two equal-amplitude sawtooth chirps to demonstrate the problem and the proposed solution. The chirps are sampled at $F_s = 20$\,MHz over a 60\,$\mu$s window with AWGN at 30\,dB SNR. Chirp~1 sweeps 1--5\,MHz with period $T_c = 20\,\mu$s and Chirp~2 sweeps 2--5\,MHz with $T_c = 40\,\mu$s. Both methods use $k_a = 0.80$ and step size $\mu = 0.35$. The LG-ANF uses $\alpha_{lp} = 0.125$ and $\alpha_P = 1.0$ whereas the standard method uses $\alpha_P = 0.15$. Figure~\ref{fig:tracking} shows the results, which are exaggerated in terms of a low $\alpha_P$ for the standard method to demonstrate the failure scenario. With the standard ANF, both stages fail to lock onto either chirp depending on the positions of the two sweeps: the notch oscillates between the two interferers' frequencies, doing nothing useful. The LG-ANF, by contrast, converges to and tracks each chirp independently in both stages (not an exclusive stage-chirp ID match though) after a short ``snapping'' phase, despite the equal-amplitude worst case.

\subsection{Amplitude-Ratio Sweep}
\label{label_simulation_amplitude_ratio}

\begin{figure}[h]
\centering
\includegraphics[width=0.99\columnwidth]{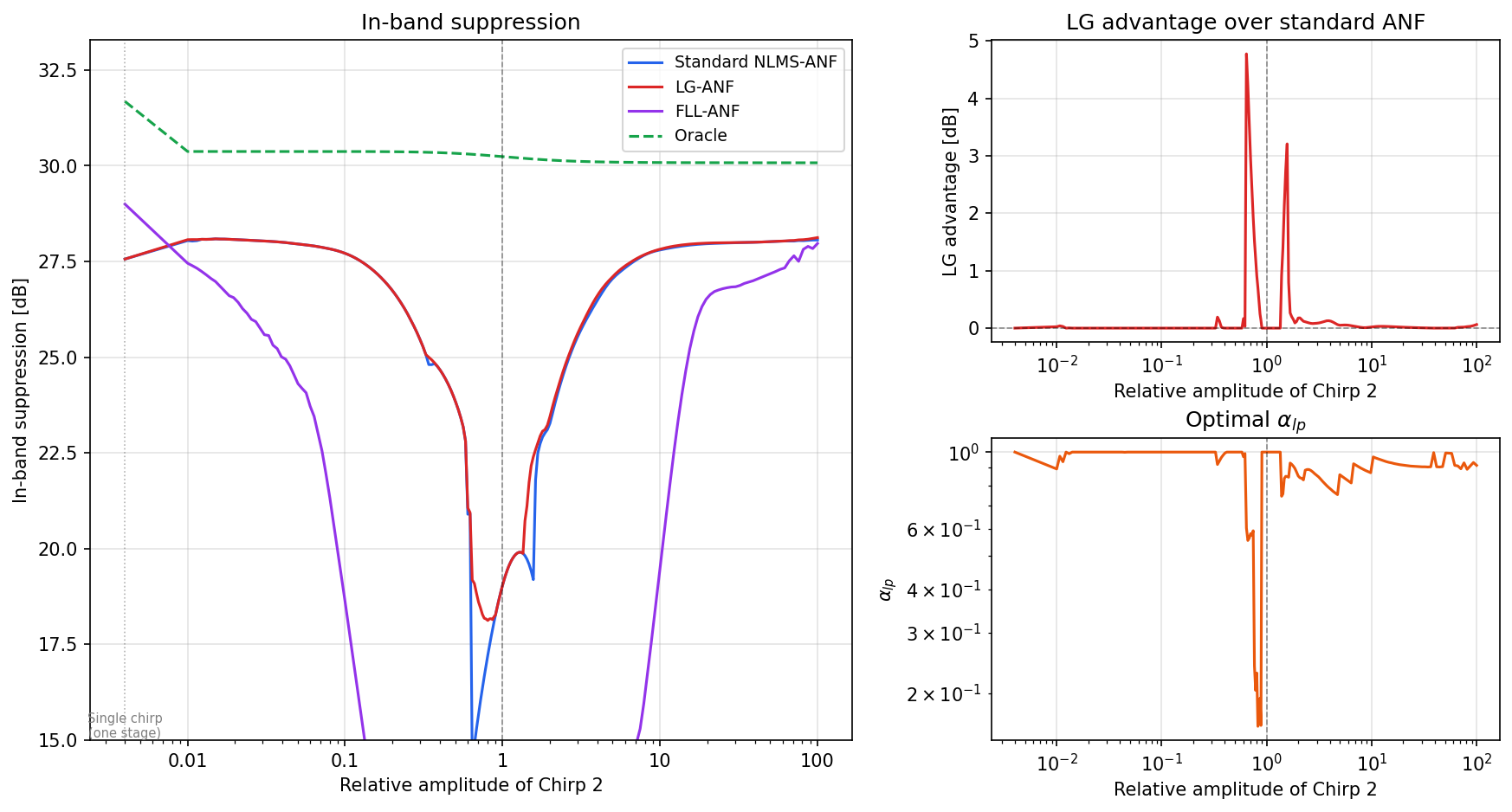}
\caption{Amplitude-ratio sweep at $\alpha_P=1$, $k_a=0.8$, $\mu=1$. Left: in-band suppression for the standard NLMS-ANF, LG-ANF, FLL-ANF, and per-sample oracle, the dotted line at the left edge marks the single-chirp ($1$-stage) point. Top-right: LG advantage over the standard ANF, peaking at two pathological ratios near equal amplitude ($\approx 4$\,dB). Bottom-right: optimal $\alpha_{lp}$ vs.\ amplitude ratio, narrowing only across the pathological band. These results represent the best achievable LG-ANF performance under per-condition optimization of $\alpha_{lp}$ and should therefore be interpreted as an upper bound on the gain obtainable from local-gradient adaptation.}
\label{fig:amp_sweep}
\end{figure}

To measure the statistical performance of the proposed LG-ANF vs.\ the NLMS-ANF, we sweep different amplitude ratios, scanning the feasible parameter ranges of $\alpha_{lp}$ and FLL loop bandwidth at each ratio, and recording the best-case values for both methods in terms of suppression performance. In the scenario, two sawtooth chirps are sampled at $F_s=2$\,MHz over a $15$\,ms window with AWGN at $30$\,dB SNR. Chirp~1 sweeps $-700$ to $+500$\,kHz and Chirp~2 sweeps $-200$ to $+800$\,kHz, both at a $100$\,MHz/s baseline rate. We compare the standard NLMS-ANF, our proposed LG-ANF, an FLL-ANF~\citep{gamba2018, vandermerwe2023}, and a per-sample oracle that notches each chirp at its true instantaneous frequency (an upper bound for this notch structure). Suppression is measured as in-band power reduction over the union of the bands of the two chirps. Figure~\ref{fig:amp_sweep} uses $\alpha_P=1$ (a controlled choice to isolate the amplitude-ratio dependence), $k_a=0.8$, and $\mu=1$, and at each amplitude ratio the best $\alpha_{lp}$ (LG) and loop bandwidth (FLL) are selected. In addition to the sweep, a single point where amplitude ratio = 0, i.e., when only a single chirp is present, is also shown. This single-point result is only to verify the advantage of FLLs in such a regime, since it's well known that FLLs perform better under single fast chirp scenarios, compared to gradient-based methods like NLMS~\citep{soner2026adam}. 

Results show that suppression degrades for both methods as the two amplitudes approach each other, but the degradation is not a single smooth dip. The degradation concentrates at specific \emph{pathological} amplitude ratios: Because the standard ANF fails through its gradient (refer to Section~\ref{label_methodology}), and the gradient is set jointly by the two chirps and the ANF parameters ($k_a$, $\mu$, $\alpha_P$), every two-chirp combination has its own pathological ratios rather than a universal worst case at ratio $1$. In the controlled sweep this appears as two sharp peaks in the LG advantage, i.e., one just below and one just above equal amplitude, where the LG-ANF recovers up to $4$\,dB of in-band suppression over the standard NLMS-ANF. Away from these points the optimal low-pass coefficient saturates to $\alpha_{lp}=1.0$ and the LG-ANF collapses to the standard NLMS update, so the two curves coincide. The optimal $\alpha_{lp}$ narrows sharply (to $\approx 0.1$) only across the pathological band. The FLL discriminator is only well posed for the single-chirp case, matching or beating the gradient methods there, but it falls well behind the gradient-based methods (standard NLMS and LG-ANF) when two chirps are present. This is expected due to the discriminator-based update rule of the FLL resembling a control loop, which degrades substantially when multiple phase components are present.

\subsection{LG Advantage Over Notch Width and Chirp Speed}
\label{label_simulation_LG_advantage}

\begin{figure*}[h]
\centering
\includegraphics[width=0.99\textwidth]{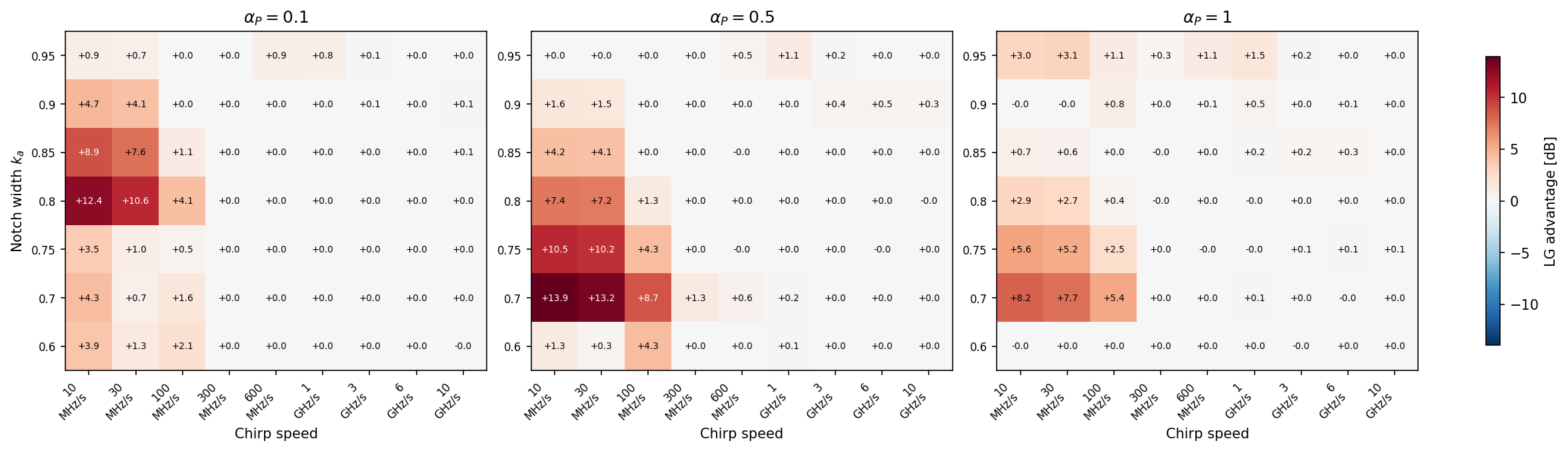}
\caption{LG advantage [dB] at the pathological amplitude ratio $0.6624$ and $\mu=1$ over notch width $k_a$ (rows) and chirp speed [Hz/s] (columns), for $\alpha_P=\{0.1,0.5,1.0\}$ (panels). The advantage reaches $13$\,dB in the most difficult parameter combinations, increases as chirps slow down, and is largest under realistic power smoothing ($\alpha_P=\{0.1$, $0.5\}$) where the EMA and local-gradient filters work together.}
\label{fig:lg_advantage_heatmap}
\end{figure*}

Figure~\ref{fig:lg_advantage_heatmap} fixes the amplitude ratio at one of the pathological values from the sweep ($0.6624$) and maps the LG advantage (LG minus standard, in dB) over notch width $k_a\in\{0.6,0.7,0.75,0.8,0.85,0.9,0.95\}$ and chirp speed from $10^7$ to $10^{10}$\,Hz/s, for three power-smoothing settings $\alpha_P\in\{0.1,0.5,1.0\}$ at fixed $\mu=1$. Because the parameter search includes $\alpha_{lp}=1$, which recovers the standard NLMS update, the optimized LG-ANF represents an upper bound on the benefit attainable through local-gradient conditioning, with gains of up to $13$\,dB for selected parameter combinations. The largest gains therefore identify conditions in which the standard gradient is particularly vulnerable, rather than representing expected gains under nominal operating conditions.

Two trends are consistent across the panels. First, the advantage decreases as the chirps get faster: rapid sweeps cross the notch dwell time before the standard gradient can be driven away from the chirp, so there is less for the local gradient to recover. Second, the dependence on the notch width $k_a$ is non-monotonic and inconclusive, although some optima exist within the swept range. Because $k_a$ cannot be chosen freely since it is tuned against the GNSS receiver impairments that the ANF itself induces, primarily the modulation-dependent group delay and timing bias of the IIR notch~\citep{borio2020}, we take the $k_a=0.9$ row as the representative operating point rather than the global maximum.

The power-smoothing constant $\alpha_P$ has a strong effect. At $\alpha_P=1.0$ the running power $P[n]$ tracks only the latest sample, so the normalized step size is very noisy and this setting is infeasible in practice. The more realistic case applies exponential moving averaging ($\alpha_P=0.1$), and the LG advantage is largest here, because the EMA power estimate and the local-gradient band-limiting work together. The LG low-pass restricts the signal that feeds the update to the local chirp band, so out-of-band noise and the competing chirp no longer corrupt the power estimate. The resulting $P[n]$ is cleaner, the normalized step size is better conditioned, and frequency tracking is more accurate than under the standard full-band update.

\section{REAL-WORLD EXPERIMENTAL EVALUATION}
\label{label_real_world_experiments}

\begin{figure}[!t]
\centering
\includegraphics[width=0.99\columnwidth]{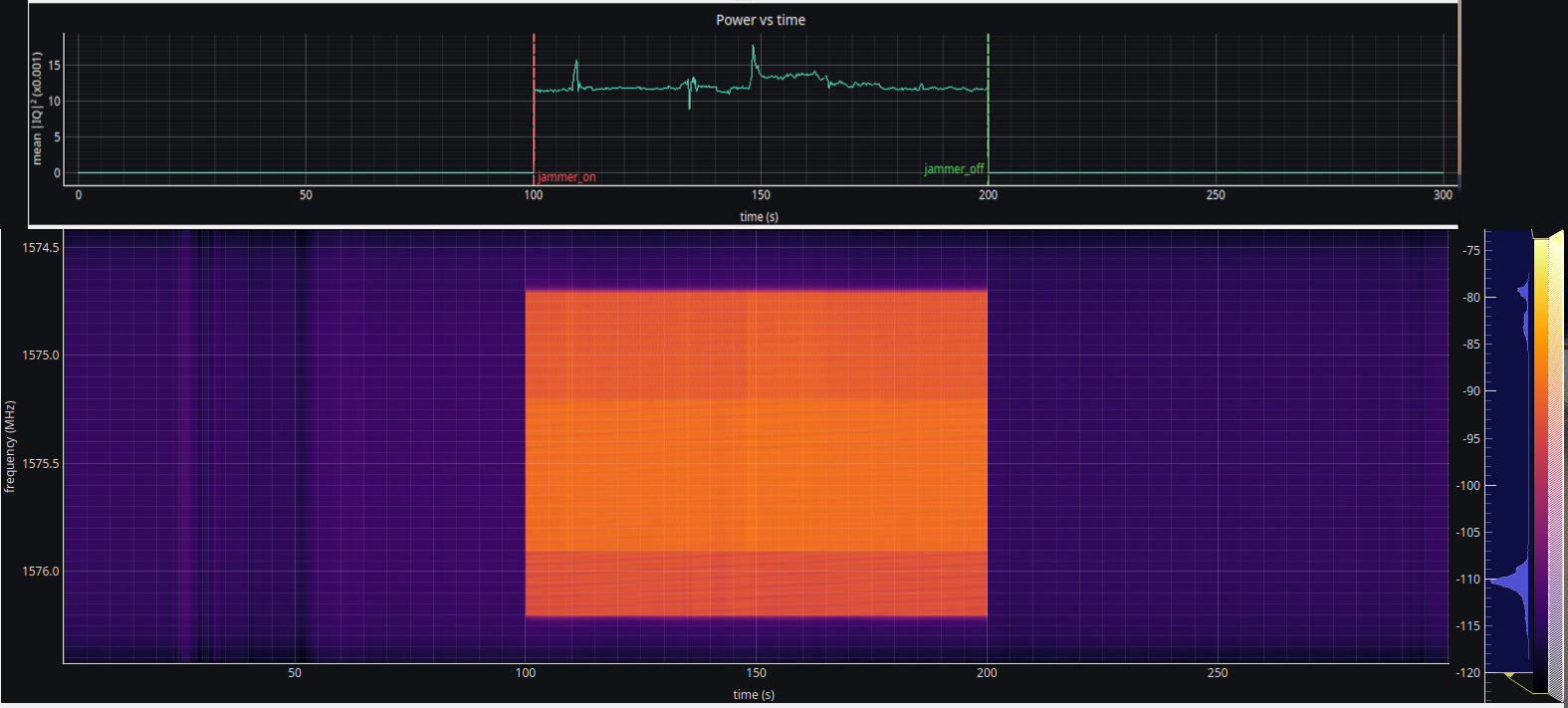}
\caption{Recorded experimental scenario. Top: mean in-phase and quadrature (IQ) power vs.\ time, with jammer-on ($100$\,s) and jammer-off ($200$\,s) markers. Bottom: spectrogram around GPS L1, the chirp jammers appear as two bright swept overlapping bands over $100$--$200$\,s.}
\label{fig:exp_psd}
\end{figure}

To verify that the simulated suppression gain translates into a downstream positioning benefit, we evaluate the LG-ANF on a recorded hardware scenario processed through a GNSS-SDR. The hardware consists of an USRP B200 receiver connected to an active GPS L1 patch antenna, and ADALM Pluto SDRs for implementing the two jammers. All experiments are done in a completely isolated space, with the necessary precautions taken to ensure that no nearby service is affected. The recording starts with a clean signal window with only GNSS signals, followed by two chirp jammers switched on at $t=100$\,s and off at $t=200$\,s around the GPS L1 center frequency. Figure~\ref{fig:exp_psd} shows the power-versus-time and spectrogram view of the recording, with the jammer-on interval clearly visible as the wideband swept energy across the L1 band. The two jammers reside on different but overlapping frequency ranges, just like in the simulation scenarios.

\subsection{Ground-Truth Reference}
\label{label_real_world_experiments_ground_truth}

A reference position is obtained from an Android smartphone in flight mode (to minimize out-of-band emissions from the phone), logging raw GNSS measurements with all available corrections. The static reference location was additionally verified against Google Maps imagery and is accurate to approximately $10$\,cm, which is well below the positioning errors of interest here, so we treat it as ground truth.

\subsection{Position-Lock Persistence}
\label{label_real_world_experimentsposition_lock}

\begin{figure}[h]
\centering
\includegraphics[width=0.87\columnwidth]{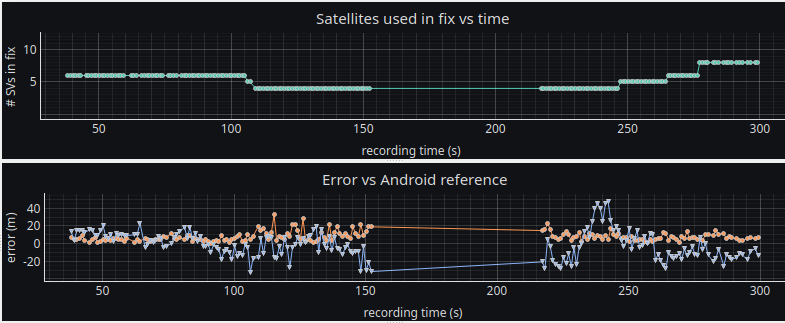}\\[2pt]
\includegraphics[width=0.87\columnwidth]{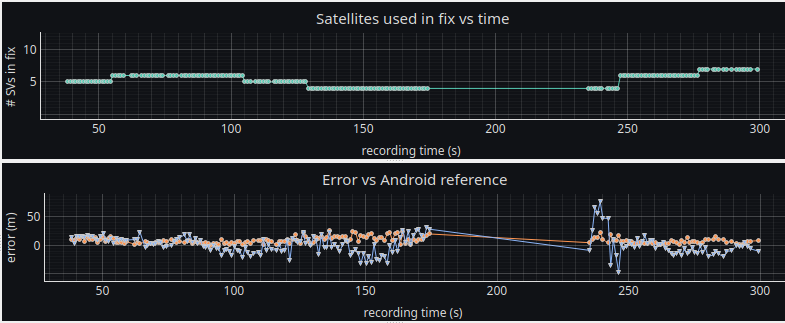}
\caption{Position-lock persistence after jammer onset at $100$\,s. Top pair: standard ANF ($\alpha_{lp}=1.0$), lock lost near $150$\,s. Bottom pair: LG-ANF ($\alpha_{lp}=0.5$), lock retained to $\approx 174$\,s. For each method: satellites used in the fix (upper) and horizontal error vs.\ the Android reference (lower).}
\label{fig:exp_lock}
\end{figure}

We process the recorded IQ with the cascaded notch over the jammer window and feed the cleaned stream to the GNSS-SDR, comparing the standard ANF ($\alpha_{lp}=1.0$) against the LG-ANF ($\alpha_{lp}=0.5$). Unlike the preceding parameter sweeps, $\alpha_{lp}=0.5$ is fixed throughout the complete recording and is not optimized as the interference conditions evolve. Figure~\ref{fig:exp_lock} reports, for each, the number of satellites used in the fix and the horizontal error against the verified reference over time. Despite the wide spread over time, the mean error over the whole recording is lower than 2 meters compared to the verified reference. After the jammer switches on at $100$\,s, the standard ANF sustains a usable position fix for about $50$\,s before losing lock at $\approx 150$\,s. The LG-ANF holds the fix for a further $24$\,s, to $\approx 174$\,s, before degrading, which shows that the additional in-band suppression provides measurable resilience in the position domain under a real chirp jammer. The post-200s recovery performance of the two differ, however that is an artifact of GNSS-SDR cold-start behavior, rather than the ANFs themselves, since the ANFs are turned off completely when the jammers stop, for a controllable test.

In addition to the lock-survival time, Figure~\ref{fig:exp_lock} also shows that the improvement is reflected in the satellite-count stability and the continuity of the position solution. After jammer onset, both methods operate close to the minimum number of satellites required for a valid fix, but the standard ANF reaches this marginal regime earlier and loses the solution near $150$\,s. The LG-ANF maintains a valid fix for a longer portion of the jammer-on interval, indicating that the local-gradient update does not merely reduce instantaneous error but helps preserve enough tracking channels for continued positioning. Therefore, the primary benefit in this experiment is not a lower average horizontal error during all valid epochs, but improved fix availability and delayed loss of lock under multi-chirp interference.

\subsection{Dataset Release}
\label{label_real_world_experiments_dataset}

To support reproducibility and further research, we release the dataset used in this evaluation, including the raw IQ recordings, the jammer and run metadata and hardware logs, and the verified reference logs, in a public GitHub repository~\citep{soner2026_dataset}. The recordings and metadata are sufficient to replicate the front-end notch processing and the GNSS-SDR results reported here.

\section{CONCLUSION}
\label{label_conclusion}

We traced the multi-chirp failure of cascaded NLMS-ANFs to gradient averaging in the full-band update, characterized the mechanism both analytically and empirically, and showed that a one-parameter local-gradient modification mitigates it. The local-gradient method collapses to the standard filter at $\alpha_{lp}=1$, making it a strict generalization of the standard NLMS approach rather than a replacement. Simulations show that each two-chirp combination has its own pathological amplitude ratios, set jointly by the gradient and the ANF parameters, where the LG-ANF recovers up to $4$\,dB of in-band suppression in a typical pathological scenario and up to $13$\,dB in edge cases, with the largest gains under realistic power smoothing of the NLMS variable step, where the power-estimation EMA filter and the local-gradient filter work synergistically. On recorded hardware data processed through the GNSS-SDR, the LG-ANF extended position-lock survival by $24$\,s beyond the standard ANF after jammer onset, confirming that the simulated suppression gain translates into a measurable positioning benefit. We release the dataset~\citep{soner2026_dataset} to enable independent reproduction and further study of the multi-chirp regime, including the $\alpha_{lp}$ activation parameters for optimal performance in real scenarios.

Several directions remain open. First, the downstream impact of the LG-ANF on GNSS signal observables, including carrier-to-noise density ($C\!/\!N_0$), pseudorange error, and carrier-phase tracking, warrants a dedicated study beyond the position-domain results reported here. Second, a systematic benchmark of the LG-ANF against the standard cascaded ANF and related methods such as the multi-parameter ANF~\citep{vandermerwe2023} and FLL-based ANFs~\citep{gamba2018}, spanning a wider range of amplitude ratios, chirp rates, frequency separations, and SNR conditions, would more fully cover the operating region in which the local-gradient correction is beneficial. Third, the improvement LG-ANF brings over NLMS-ANF is limited (up to 4 dB under practical choice of parameters), motivating further studies on the topic. Finally, since the LG-ANF only departs from the standard update across the pathological band, a lightweight activation criterion for transparent switching between standard and LG modes would preserve single-interferer and dissimilar-amplitude performance while engaging the local gradient only when needed, yielding a robust, deployable upgrade for GNSS receivers facing emerging multi-PPD interference.

\bibliographystyle{apalike}
\bibliography{ION_GNSS}

\end{document}